\documentclass[11pt]{article}
\usepackage[preprint]{acl}
\usepackage{times}
\usepackage{latexsym}
\usepackage[T1]{fontenc}
\usepackage[utf8]{inputenc}
\usepackage{microtype}
\usepackage{inconsolata}
\usepackage{graphicx}
\usepackage{amsmath}
\usepackage{amssymb}
\usepackage{tcolorbox}
\usepackage{cleveref}
\usepackage{booktabs}
\usepackage{colortbl}
\usepackage{enumitem}

\newtcolorbox{researchquestions}{
    colback=blue!7!white,
    colframe=blue!75!black,
    fontupper=\normalsize,
    boxrule=1pt,
    arc=2mm,
    left=7pt,right=7pt,top=6pt,bottom=6pt,
    before skip=6pt,after skip=8pt
}

\title{QProtFuse: Query-Guided Layer Fusion for Retrieval-Augmented Protein Function Prediction}
{\footnotesize
\author{
\textbf{Mianzhi Pan\textsuperscript{1,2}}\quad
\textbf{Peishuo Liu\textsuperscript{1}}\quad
\textbf{Yawen Ouyang\textsuperscript{3}}\quad
\textbf{Botian Wang\textsuperscript{2}}\\
\textbf{Yiming Rong\textsuperscript{5}}\quad
\textbf{Hao Zhou\textsuperscript{2}}\thanks{Corresponding author}\quad
\textbf{Jianbing Zhang\textsuperscript{1,4}}\footnotemark[\value{footnote}]\\[0.075em]
\textsuperscript{1}National Key Laboratory for Novel Software Technology \&\\
School of Artificial Intelligence, Nanjing University, China\\
\textsuperscript{2}Institute of AI Industry Research (AIR), Tsinghua University\quad
\textsuperscript{3}Shanghai AI Lab\\
\textsuperscript{4}Chemistry and Biomedicine Innovation Center (ChemBIC), Nanjing University\\
\textsuperscript{5}University of Chinese Academy of Sciences, Beijing\\
}}

\begin{document}
\maketitle
\begin{abstract}
Predicting protein functions from amino acid sequences is essential to understanding biological processes. Large language models (LLMs) offer flexible function descriptions to natural-language queries, yet their limited protein understanding motivates retrieval-augmented generation (RAG).
Existing protein RAG methods primarily expand external retrieval signals while overlooking the rich knowledge encoded within a single protein language model (PLM).
We show that retrieval performance varies across PLM layers and biological tasks, making conventional final-layer selection suboptimal. 
To this end, we propose QProtFuse, which uses queries to adaptively fuse multi-layer protein features and jointly learns protein and annotation retrieval through intra-modal and cross-modal contrastive objectives. 
On Prot-Inst-OOD, QProtFuse achieves the best protein and annotation retrieval scores, reaching 16.4\% Recall@1 with annotation reranking. These gains improve downstream RAG-based function descriptions, yielding the highest LLM-as-a-judge scores for both Qwen3-8B (6.22/10) and DeepSeek-V4-Pro (6.68/10).
Ablations and weight analyses demonstrate the value of query-guided fusion and complementary training objectives for exploiting a PLM's internal knowledge.\footnote{Data and code are available at \url{https://github.com/panmianzhi/QProtFuse}.}
\end{abstract}

\section{Introduction}
\label{sec:introduction}

Proteins are essential to biological processes. Predicting their functions from amino acid sequences is therefore a central task in computational biology. Traditional sequence analysis uses BLAST~\citep{altschul1990blast} to find homologs and InterPro~\citep{blum2025interpro} to identify family and domain signatures. These methods depend on informative matches and cannot directly answer specific biological questions. More recently, broader large language models (LLMs) for life-science research have begun to tackle protein-related tasks~\citep{openai2026rosalind,anthropic2026lifesciences}. Protein-oriented LLMs incorporate amino acid sequences through direct tokenization~\citep{wang-etal-2024-instructprotein,bai2025interns1,qiu2026amix2} or dedicated protein encoders~\citep{zhuo-etal-2024-protllm,fei2025prot2text,zhou2026evolla}. Together, these models offer the potential to answer complex biological questions in free-form by leveraging their internal pretrained knowledge and reasoning capabilities.

\begin{figure}[t]
    \centering
    \includegraphics[width=\columnwidth]{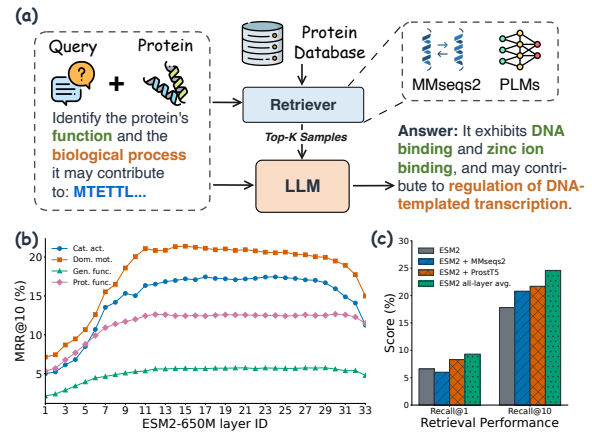}
    \caption{
    \textbf{(a)} A typical retrieval-augmented pipeline for protein function description.
    \textbf{(b)} Layer-wise retrieval performance of ESM2-650M across four Prot-Inst-OOD tasks.
    \textbf{(c)} Retrieval performance comparison of final-layer ESM, ESM with MMseqs2, ESM with ProstT5, and averaged ESM layer features.}
    \label{fig:motivation}
\end{figure}

Despite these advances, the inherent gap between proteins and natural language remains a major obstacle to reliable protein understanding by LLMs~\citep{zhuang2026lost}. For protein function description, \citet{wu-etal-2025-rethinking-text} show that retrieval-based methods can outperform direct LLM generation. This highlights the value of external biological knowledge, motivating retrieval-augmented generation (RAG) approaches~\citep{wu-etal-2025-rethinking-text,ma2026protquest,ding2026proteinrag}. As illustrated in \Cref{fig:motivation}(a), these approaches retrieve relevant proteins and annotations to enhance LLM-based protein function description.

However, current protein RAG methods focus on incorporating external retrieval signals, such as those from sequence alignment tools~\citep{altschul1990blast,steinegger2017mmseqs2} and additional protein language models (PLMs)~\citep{lin2023esmfold,wang2024dplm,heinzinger2024bilingual}, while leaving the knowledge within a single PLM insufficiently explored. Previous studies reveal that PLMs encode rich and distinct biological features across different layers~\citep{rao2021structure,lin2023esmfold,vig2021bertology,kumar2025layer}. These findings raise two questions:
\begin{researchquestions}
\textbf{Q1:} Do different PLM layers exhibit distinct strengths in protein retrieval?\par\smallskip
\textbf{Q2:} If so, how can we effectively use this internal knowledge to improve protein RAG?
\end{researchquestions}

To answer these questions, we first probe the layerwise retrieval performance of ESM-2~\citep{lin2023esmfold} on Prot-Inst-OOD~\citep{wu-etal-2025-rethinking-text}. As shown in \Cref{fig:motivation}(b), the four subtasks achieve their best retrieval performance using features from different intermediate layers. These observations answer Q1 and show that relying solely on final-layer features is suboptimal. 
As a first step toward Q2, we test a simple fusion strategy by uniformly averaging features across all layers. Even this simple strategy outperforms those incorporating MMseqs2 sequence similarity or ProstT5 features (\Cref{fig:motivation}(c)).\footnote{For the two compared methods, we follow the standard retrieval designs of RAPM~\citep{wu-etal-2025-rethinking-text} and the structure-only variant of ProtQueSt~\citep{ma2026protquest}, respectively.} 
However, uniform averaging still generally falls short of task-specific best-performing layers (see Appendix~\ref{sec:appendix_layer_comparison}). This gap calls for a more adaptive fusion strategy that exploits the distinct strengths of different layers.

Inspired by multi-layer visual feature fusion in multimodal LLMs~\citep{yao2024dense,lin2025visualfusion,lin2026tgif}, we propose \textbf{QProtFuse} (Query-guided multi-layer Protein Feature Fusion) for retrieval-augmented protein function prediction. 
QProtFuse leverages the query to \textbf{adaptively assign layer-wise weights} and fuse multi-layer features.
To optimize this fusion, we propose a \textbf{dual-objective contrastive learning} framework that couples an intra-modal protein-to-protein objective with an inter-modal protein-to-annotation objective.
These two objectives are mutually reinforcing and jointly enable both protein and annotation retrieval during inference, thereby broadening the applicability of QProtFuse.
On Prot-Inst-OOD, QProtFuse achieves \textbf{superior protein and annotation retrieval performance} compared with all baselines, reaching a Recall@1 of 16.4\% and Recall@10 of 25.9\% with our reranking configuration. 
These retrieval gains improve downstream function descriptions, yielding the \textbf{highest LLM-as-a-judge scores} among the evaluated methods, with 6.22/10 on Qwen3-8B and 6.68/10 on DeepSeek-V4-Pro.

\noindent Our contributions are summarized as follows:
\begin{itemize}[leftmargin=*, itemsep=0.5pt, parsep=0pt, topsep=5pt]
    \item We present the first systematic study of multi-layer PLM features for protein retrieval, showing that the best layer varies across tasks and using only final-layer features is suboptimal.
    \item We propose QProtFuse, which combines query-guided layer fusion with joint protein-to-protein and protein-to-annotation alignment to exploit the internal knowledge of PLMs.
    \item We achieve SOTA retrieval and RAG-based protein function description performance, demonstrating the value of adaptive layer fusion for protein search and annotation.
\end{itemize}

\section{Related Work}
\label{sec:related_work}

\subsection{LLMs for Protein Understanding}
\label{sec:related_protein_llms}

\begin{figure*}[t]
    \centering
    \includegraphics[width=.95\textwidth]{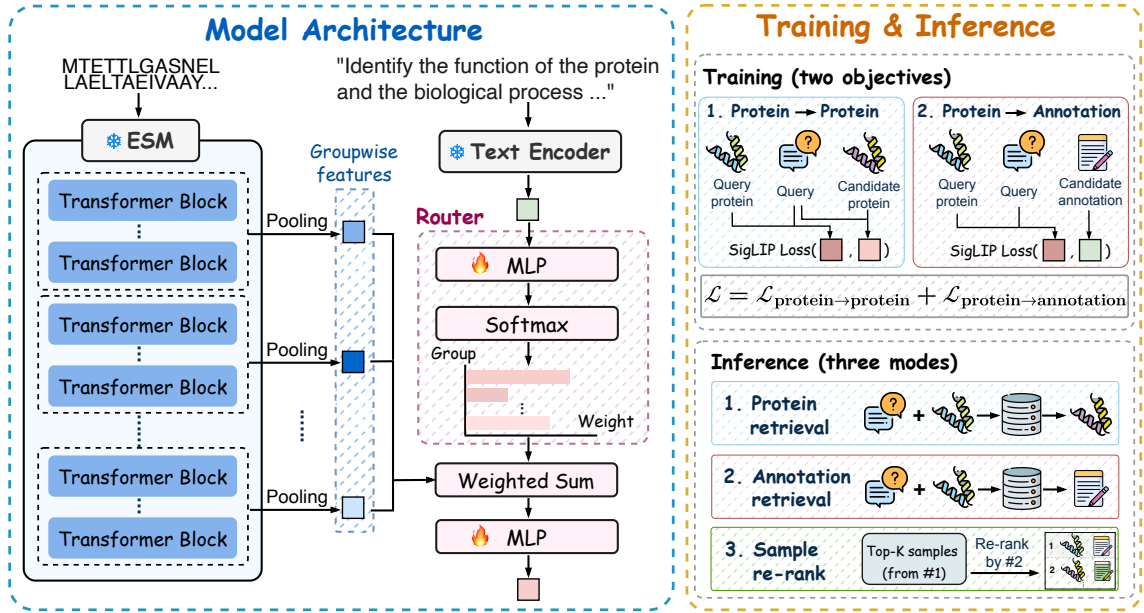}
    \caption{Overview of QProtFuse Framework. \textbf{Left.} Features from a frozen ESM are pooled within consecutive layer groups and fused using query-dependent weights. \textbf{Right.} During training, the model is jointly optimized for intra-modal protein-protein and inter-modal protein-to-annotation alignment. During inference, one checkpoint supports protein retrieval, annotation retrieval, and protein retrieval with annotation-based reranking.}
    \label{fig:architecture}
\end{figure*}

Protein-oriented LLMs mainly integrate proteins through direct tokenization or encoder-based alignment. Direct tokenization approaches model amino-acid tokens alongside natural language~\citep{wang-etal-2024-instructprotein,xia2025nature}, with further developments in both tokenization and training objectives, where Intern-S1~\citep{bai2025interns1} introduces modality-specific tokenization and AMix-2~\citep{qiu2026amix2} adopts block-wise diffusion for joint protein--text modeling. Encoder-based approaches instead use projectors or Q-Formers to align protein features with an LLM~\citep{zhuo-etal-2024-protllm,liu2024prott3,zhou2026evolla,wang2025prot2chat}. Within this paradigm, Prot2Text-V2~\citep{fei2025prot2text} strengthens protein--text alignment through contrastive learning, while BioReason-Pro~\citep{fallahpour2026bioreason} emphasizes biological reasoning through reinforcement learning. Despite these advances, \citet{zhuang2026lost} identify limitations in both approaches, motivating an alternative that supplies LLMs with structured biological context extracted by external tools.

Reliable protein function description remains challenging for LLMs due to the inherent modality gap. \citet{wu-etal-2025-rethinking-text} show that simple retrieval can outperform protein-oriented LLMs and propose a dual-index retrieval method combining MMseqs2~\citep{steinegger2017mmseqs2} and ESM~\citep{lin2023esmfold}. 
ProtQueSt~\citep{ma2026protquest} ensembles ESM and ProstT5~\citep{heinzinger2024bilingual} features with query-conditioned contrastive learning to retrieve annotations suited to the biological question. These methods improve retrieval through more external similarity signals. Our work focuses on the knowledge within a single PLM, learning to combine its layer features according to the query.

\subsection{Layerwise Probing in PLMs}
\label{sec:related_plm_layers}

Recent studies have probed the biological knowledge encoded within PLMs. \citet{rao2021structure} and \citet{lin2023esmfold} show that PLMs learn structural information from sequences alone that supports residue contact prediction and atomic-level folding. 
More specifically, \citet{vig2021bertology} find that attention captures residue contacts and binding sites, with deeper layers attending to increasingly complex biophysical properties. \citet{kumar2025layer} probe ESM-2 layers for kinase functional prediction and show that selected intermediate layers outperform the final layer. 
From a geometric perspective, \citet{beshkov2026towards} trace changes in feature geometry across depth and find that structural fidelity is highest near, but before, the final layer. 
Together, these studies establish that the most useful biological features need not occur at the encoder output. We investigate how these differences affect protein retrieval and how query-guided fusion can turn them into more effective retrieval features.

\subsection{Multi-Layer Feature Fusion}
\label{sec:related_layer_fusion}

Motivated by the observation that earlier visual layers preserve visual details that are not fully captured by final layers, recent work has studied multi-layer fusion in multimodal LLMs. Dense Connector~\citep{yao2024dense} explores integrating features from multiple visual encoder layers to enrich visual input.
\citet{chen-etal-2025-multimodal} show that shallow and middle layers are particularly useful for fine-grained visual tasks and develop a lightweight layer-concat fusion strategy. 
\citet{li2026instruction} and \citet{lin2026tgif} use textual queries to adaptively determine the weights of visual features from different layers. \citet{lin2025visualfusion} conduct a systematic comparison and find that combining features from different encoder layers improves generalization and directly fusing visual features at the LLM input, rather than injecting them into its intermediate layers, provides consistently strong performance.
We extend layer fusion to PLMs and adapt it to retrieval through joint protein-to-protein and protein-to-description contrastive objectives.

\section{Method}
\label{sec:method}

We study retrieval-augmented protein function description given natural-language queries. Each sample is a tuple $(p,q,d)$ comprising a protein sequence $p$, a natural-language query $q$, and its functional description $d$. Given a reference database $\mathcal{D}=\{(p_i,q_i,d_i)\}_{i=1}^{M}$, retrieval-augmented prediction selects $K$ reference samples $\mathcal{S}_K$ to enhance LLM generation of $d$. For a fixed LLM generator $P_\phi$, the ideal evidence set satisfies
\begin{equation}
    \mathcal{S}_K^*=\underset{\mathcal{S}\subseteq\mathcal{D},\,|\mathcal{S}|=K}{\arg\max}
    \log P_\phi(d\mid p,q,\mathcal{S}).
    \label{eq:evidence_objective}
\end{equation}
Our goal is to train a retriever that selects such evidence. However, directly optimizing \Cref{eq:evidence_objective} is expensive, as the $\binom{M}{K}$ possible combinations form a prohibitively large search space, and the evaluation requires costly LLM inference. 
Prior work shows that improving retrieved evidence can enhance generation quality~\citep{lewis2020rag,shi-etal-2024-replug}. We therefore focus on retrieval effectiveness as a proxy for evidence quality:
\begin{equation}
    \theta^*=\underset{\theta}{\arg\max}\;
    \mathbb{E}_{(p,q,d)\sim\mathcal{T}}\!\left[
    \mathcal{U}\bigl(\pi_\theta(p,q),\mathcal{R}(d)\bigr)
    \right],
    \label{eq:retrieval_selection}
\end{equation}
where $\mathcal{T}$ is the training set, $\pi_\theta(p,q)$ ranks the samples in $\mathcal{D}$ by $(p,q)$, and $\mathcal{R}(d)$ is the set of oracle samples in $\mathcal{D}$ that match description $d$. 
$\mathcal{U}$ measures ranking quality, such as Recall@$K$.

\Cref{fig:architecture} illustrates the QProtFuse framework. 
We evenly divide consecutive PLM layers into groups and mean-pool their features, then use the query to predict group weights for adaptive fusion (\Cref{sec:architecture}). We train QProtFuse with two complementary objectives: intra-modal protein-to-protein (PP) alignment and cross-modal protein-to-annotation (PA) alignment (\Cref{sec:objectives}). The same model supports protein retrieval, annotation retrieval, and two-stage retrieval with reranking (\Cref{sec:retrieval_modes}). 

\subsection{Query-Guided Layer Fusion}
\label{sec:architecture}

\begin{figure}[t]
    \centering
    \includegraphics[width=.85\columnwidth]{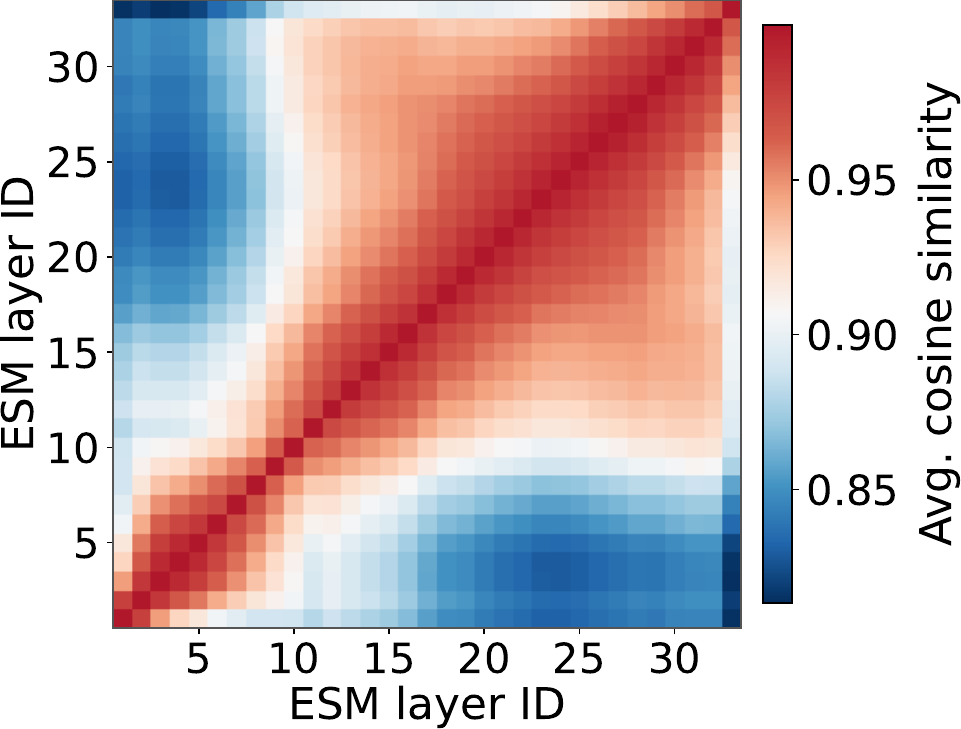}
    \caption{Pairwise cosine similarity between ESM-2 layer features, averaged over training proteins. Neighboring layers typically produce similar features, motivating consecutive-layer grouping.}
    \label{fig:layer_similarity}
\end{figure}

\paragraph{Grouping layer features.}
For a protein $p$ with $n$ residues, a frozen ESM2-650M encoder~\citep{lin2023esmfold} produces residue features $\mathbf{H}^{(\ell)}(p)\in\mathbb{R}^{n\times1280}$ at each of its $L=33$ transformer layers. We average over residues to obtain a protein feature for each layer. 
As shown in \Cref{fig:layer_similarity}, neighboring ESM layers typically yield highly similar protein features, suggesting redundancy in routing over every layer independently. 
We therefore follow the consecutive-layer grouping strategy of \citet{li2026instruction}, and partition the layers into $G=7$ groups $\{\mathcal{G}_g\}_{g=1}^{G}$ and average within each group:
\begin{equation}
    \begin{aligned}
    \mathbf{l}_\ell(p)&=\frac{1}{n}\sum_{t=1}^{n}\mathbf{H}^{(\ell)}_t(p),\\
    \mathbf{g}_g(p)&=\frac{1}{|\mathcal{G}_g|}\sum_{\ell\in\mathcal{G}_g}\mathbf{l}_\ell(p).
    \end{aligned}
    \label{eq:group_features}
\end{equation}
Grouping reduces the number of fusion weights while retaining features from different depths.

\paragraph{Query-guided fusion.}
As shown in prior studies~\citep{vig2021bertology,beshkov2026towards}, different PLM layers capture distinct biological properties with varying relevance to downstream tasks. We therefore condition layer fusion on the query, adapting each group's contribution to the requested functional aspect. A frozen text encoder $T$ maps $q$ to $\mathbf{e}_q=T(q)$, from which an MLP predicts group weights:
\begin{equation}
    \boldsymbol{\alpha}(q)=\operatorname{softmax}\!\left(
    \operatorname{MLP}_q(\mathbf{e}_q)
    \right).
    \label{eq:group_weights}
\end{equation}
Using these weights, we fuse the group features to obtain the final query-conditioned protein feature:
\begin{equation}
    \begin{aligned}
    \mathbf{f}(p\mid q)&=\sum_{g=1}^{G}\alpha_g(q)\mathbf{g}_g(p),\\
    \mathbf{z}(p\mid q)&=\operatorname{L2Norm}\!\left(\operatorname{MLP}(\mathbf{f}(p\mid q))\right).
    \end{aligned}
    \label{eq:protein_projection}
\end{equation}

\paragraph{A high-level view.}
QProtFuse makes the features of a protein conditional on the biological question. From a broader perspective, this follows the principle of conditional representation learning, which adapts features to a specified task~\citep{liu2026conditional}, and is consistent with the query-conditioned protein retrieval formulation of ProtQueSt~\citep{ma2026protquest}. Our conditioning acts on the mixture of PLM layers, where the same sequence can receive different combinations of features from a frozen PLM under different queries.

\subsection{Joint Contrastive Objectives}
\label{sec:objectives}

A natural way to construct $\pi_\theta(p,q)$ in \Cref{eq:retrieval_selection} is to use $\mathbf{z}(p\mid q)$ from \Cref{eq:protein_projection} to rank candidates according to their descriptions. This leads to protein--text cross-modal contrastive learning, as used by \citet{ma2026protquest}. However, cross-modal alignment alone can neglect within-modality relationships and degrade feature quality~\citep{yang2022vision}. We therefore combine intra-modal PP and cross-modal PA alignment objectives to encourage query-conditioned similarity between proteins while aligning them with their annotations.

\paragraph{Protein-to-protein alignment.}
Given positive and negative pairs with labels $y_{ij}\in\{+1,-1\}$, the PP objective learns to distinguish functionally related proteins under the query. We encode the query protein and each candidate with the same instruction $q_i$ and compute their similarity:
\begin{equation}
    r^{\mathrm{PP}}_{ij}=\cos\!\left(\mathbf{z}(p_i\mid q_i),\mathbf{z}(p_j\mid q_i)\right)/\tau.
    \label{eq:pp_score}
\end{equation}
We optimize the expected SigLIP loss~\citep{zhai2023sigmoid} over training queries $s_i\in\mathcal{T}$ and candidates $s_j\in\mathcal{D}$:
\begin{equation}
    \mathcal{L}_{\mathrm{PP}}=
    \mathbb{E}_{s_i\sim\mathcal{T},\,s_j\sim\mathcal{D}}\!\left[
    -\log\sigma\!\left(y_{ij}\bigl(r^{\mathrm{PP}}_{ij}+b\bigr)\right)
    \right].
    \label{eq:pp_loss}
\end{equation}
Here $\sigma$ is the sigmoid function, and both the temperature $\tau$ and bias $b$ are learnable. During training, $\mathcal{D}=\mathcal{T}$, with candidate selection specified below. 
This objective increases query-conditioned similarity for positive protein pairs and decreases it for negative pairs.

\paragraph{Protein-to-annotation alignment.}
The PA objective aligns the same query-conditioned protein features with candidate annotation features $\mathbf{u}(d_j)$, obtained from the same frozen text encoder $T$ used for queries and a separate annotation projector. We define the similarity and expected loss as
\begin{equation}
    \begin{aligned}
    r^{\mathrm{PA}}_{ij}&=\cos\!\left(\mathbf{z}(p_i\mid q_i),\mathbf{u}(d_j)\right)/\tau,\\
    \mathcal{L}_{\mathrm{PA}}&=
    \mathbb{E}_{s_i\sim\mathcal{T},\,s_j\sim\mathcal{D}}\!\left[
    -\log\sigma\!\left(y_{ij}\bigl(r^{\mathrm{PA}}_{ij}+b\bigr)\right)
    \right].
    \end{aligned}
    \label{eq:pt_loss}
\end{equation}
This objective draws protein features toward matching annotations and pushes them from unrelated descriptions. We combine the two objectives into the final objective:
\begin{equation}
    \mathcal{L}=\mathcal{L}_{\mathrm{PP}}+\mathcal{L}_{\mathrm{PA}}.
    \label{eq:joint_loss}
\end{equation}
Both objectives update the fusion network. The ESM and text encoder remain frozen.

\paragraph{Positive and negative pairs.}
For a query sample $s_i=(p_i,q_i,d_i)$, positives share its target description, while eligible negatives have similar instructions but dissimilar descriptions. We adopt the setting of ProtQueSt~\citep{ma2026protquest} and define
\begin{equation}
    \begin{aligned}
    \mathcal{P}_i&=\{s_j\in\mathcal{D}:\cos(q_i,q_j)>\tau_{\text{query}},d_j=d_i\},\\
    \mathcal{N}_i&=\{s_j\in\mathcal{D}:\cos(q_i,q_j)>\tau_{\text{query}},\\
    &\hspace{67pt}\cos(d_i,d_j)<\tau_{\text{fn}}\}.
    \end{aligned}
    \label{eq:positive_negative_pairs}
\end{equation}
$\tau_{\text{query}}$ restricts both positive and negative pairs to samples with similar instructions, while $\tau_{\text{fn}}$ filters out potential false negatives with similar annotations.
From $\mathcal{N}_i$, we select up to $H$ hard negatives with the highest similarity under all-layer-mean ESM features and upweight them during training. Positive pairs receive $y_{ij}=+1$ and negative pairs receive $y_{ij}=-1$.

\subsection{Inference}
\label{sec:retrieval_modes}

Joint training enables QProtFuse to retrieve evidence for RAG through three complementary modes: protein retrieval, annotation retrieval, and protein retrieval with annotation reranking. All three modes use the same checkpoint.

\paragraph{Protein retrieval.}
The PP branch performs sequence-based protein retrieval using \Cref{eq:pp_score}. 
The same protein can yield different rankings when queried with respect to distinct biological perspectives, such as catalytic activity or domain organization.
This makes protein retrieval responsive to the biological question, while retaining its applicability to candidates with missing, incomplete, or unreliable annotations.

\paragraph{Annotation retrieval.}
The PA branch supports cross-modal search from a protein and its query to functional descriptions using $r^{\mathrm{PA}}$ in \Cref{eq:pt_loss}. Existing methods~\citep{su2025trimodal,zhou2025protclip} align text with proteins whose features are independent of functional queries. 
QProtFuse, in contrast, conditions the protein features on $q$, allowing annotation relevance to vary with the requested functional aspect. This enables direct retrieval of textual evidence tailored to the query.

\paragraph{Protein retrieval with annotation reranking.}
For databases containing paired proteins and annotations, QProtFuse enables us to retrieve a shortlist by using PP to identify proteins from query-conditioned features, and then rerank its entries using PA scores, which refine the order based on how well annotations match the query protein and question. This reranking step can improve RAG by prioritizing relevant evidence within an initial candidate set~\citep{yu2024rankrag}, thereby surfacing informative annotations for downstream generation.

\section{Experiments}
\label{sec:experiments}

In this section, we investigate the value of exploiting a single PLM's multi-layer features for retrieval-augmented protein function prediction. Through baseline comparisons and controlled ablations, we address three questions: (1) Can QProtFuse improve protein retrieval? (2) How do query-guided fusion and the PP and PA objectives contribute? (3) Do retrieval improvements translate into better RAG-generated function descriptions?

\subsection{Experimental Setup}
\label{sec:experimental_setup}

\paragraph{Dataset.}
We use Prot-Inst-OOD~\citep{wu-etal-2025-rethinking-text}, an instruction-based protein understanding benchmark covering protein function, catalytic activity, domain/motif, and general function. 
It uses an out-of-distribution split to reduce annotation leakage between training and test sets. 
We train QProtFuse on the training split. 
Evaluation follows official test-to-train retrieval: all 14,503 test samples serve as queries, and the 284,526 training records from all four tasks form a single shared reference database. Appendix~\ref{sec:appendix_dataset} provides task descriptions, split construction, and dataset statistics.

\paragraph{Metrics.}
We report Recall@$K$ (R@$K$) and Mean Reciprocal Rank (MRR) for retrieval. 
For generated descriptions, we report the standard NLP metrics ROUGE-L~\cite{lin2004rouge} for lexical overlap and Entity-BLEU~\citep{wu-etal-2025-rethinking-text}, a biological entity-focused metric which computes the standard BLEU score~\citep{papineni2002bleu} on extracted biological entities.
We also adopt ProtQueSt's LLM-as-a-judge metric~\citep{ma2026protquest}, assessing biological recall, precision, specificity, and plausibility. Scores are computed per task and macro-averaged. 
Definitions and calculations of these metrics are provided in Appendix~\ref{sec:appendix_generation_metrics}.

\paragraph{Baselines.}
We compare QProtFuse with two representative RAG-based function description methods: RAPM~\citep{wu-etal-2025-rethinking-text} and ProtQueSt~\citep{ma2026protquest}. 
For retrieval, we additionally include ESM all-layer mean and the sequence-based retriever MMseqs2 for protein-to-protein retrieval, and ProTrek~\citep{su2025trimodal} for protein-to-text retrieval.
For generation, we also compare general-purpose LLMs without RAG and protein-oriented LLMs. 
All RAG-based methods use Top-10 retrieved references per query for generation.
Generation details are provided in Appendix~\ref{sec:appendix_generation_prompts}.

\begin{table}[t]
    \centering
    \fontsize{10pt}{10.5pt}\selectfont
    \setlength{\tabcolsep}{3.5pt}
    \caption{Retrieval results on Prot-Inst-OOD (\%). We average QProtFuse scores over three seeds. ``$^{*}$'' denotes the two-channel ``5+5'' strategy of ProtQueSt, with no instruction filtering during the structural channel. Both variants marked ``rerank'' use our reranking strategy.}
    \label{tab:retrieval_results}
    \begin{tabular}{@{}lccc@{}}
        \toprule
        Method & MRR@10 & R@1 & R@10 \\
        \midrule
        \rowcolor{gray!15}
        \multicolumn{4}{@{}l@{}}{\textbf{Protein to protein}} \\
        MMseqs2 & 9.5 & 5.5 & 19.0 \\
        ESM2-650M$_\text{all-layer avg.}$ & 14.0 & 9.3 & 24.6 \\
        RAPM & 11.0 & 6.0 & 20.7 \\
        ProtQueSt$_\text{struct-only}$ & 12.5 & 8.3 & 21.7 \\
        QProtFuse$_\text{pp}$ & $\mathbf{14.7}$ & $\mathbf{9.7}$ & $\mathbf{25.9}$ \\
        \midrule
        \rowcolor{gray!15}
        \multicolumn{4}{@{}l@{}}{\textbf{Protein to annotation}} \\
        ProTrek & 12.3 & 7.9 & 22.7 \\
        ProtQueSt$_\text{FiLM-only}$ & 17.5 & 12.4 & 28.2 \\
        QProtFuse$_\text{pa}$ & $\mathbf{20.4}$ & $\mathbf{15.5}$ & $\mathbf{29.2}$ \\
        \midrule
        \rowcolor{gray!15}
        \multicolumn{4}{@{}l@{}}{\textbf{Hybrid}} \\
        ProtQueSt$^{*}$ & 8.2 & 3.0 & \textbf{26.8} \\
        ProtQueSt$_\text{rerank}$ & 17.1 & 15.0 & 21.7 \\
        QProtFuse$_\text{rerank}$ & $\mathbf{18.6}$ & $\mathbf{16.4}$ & $25.9$ \\
        \bottomrule
    \end{tabular}
\end{table}

\paragraph{Implementation details.}
QProtFuse uses OpenAI's \texttt{text-embedding-3-large} (3072-D) for query and annotation embedding. 
We set $\tau_{\text{query}}=0.65$, $\tau_{\text{fn}}=0.70$, and $H=128$, and initialize the learnable SigLIP temperature $\tau$ and bias $b$ to $0.07$ and $-10$, respectively. 
Further training hyperparameters are provided in Appendix~\ref{sec:appendix_training}. We apply instruction-based candidate filtering with a similarity threshold of $0.65$ across retrieval methods.\footnote{For ProtQueSt, we follow its published ``5+5'' setting and omit instruction filtering during structural retrieval.}
We apply this filter because each record is annotated only for the functional aspect specified by its instruction. Otherwise, a relevant protein could be treated as a false negative simply because its recorded answer addresses a different question.

\subsection{Results}
\label{sec:results}

\subsubsection{Retrieval Results}
\label{sec:retrieval_results}
\Cref{tab:retrieval_results} shows that QProtFuse achieves the highest scores on all three metrics within both PP and PA retrieval. Compared with ESM all-layer averaging, QProtFuse improves MRR from 14.0 to 14.7 and R@10 from 24.6 to 25.9, demonstrating gains beyond uniform fusion of the PLM features. 
QProtFuse also consistently outperforms RAPM and ProtQueSt in PP retrieval, as well as ProtQueSt in PA retrieval. These gains over methods integrating additional external signals highlight the value of exploiting the rich inner knowledge within a single PLM.
Adding annotation reranking to PP retrieval further improves performance for both ProtQueSt and QProtFuse, indicating that relevant samples are ranked higher, which can benefit downstream RAG~\citep{yu2024rankrag}. 
QProtFuse outperforms reranked ProtQueSt on all three metrics and achieves the highest MRR and R@1 among the Hybrid methods. We use this reranking configuration for downstream RAG.
Note that PA ranks only 59,065 candidate annotations, whereas PP and Hybrid retrieve from the full training set ($\sim$284K). Their scores are therefore not directly comparable.

\subsubsection{RAG Generation Results}
\label{sec:rag_results}
\begin{table}[t]
    \centering
    \fontsize{10pt}{11pt}\selectfont
    \setlength{\tabcolsep}{8pt}
    \caption{Generation results on Prot-Inst-OOD. Prot2Text-V2~\citep{fei2025prot2text} is trained on the training split and serves as a supervised comparison. E-BLEU is Entity-BLEU and RG-L is ROUGE-L (both in \%). Judge-F is the final judge score (0--10).}
    \label{tab:generation_results}
    \begin{tabular}{@{}lccc@{}}
        \toprule
        Method & E-BLEU & RG-L & Judge-F \\
        \midrule
        \rowcolor{gray!15}
        \multicolumn{4}{@{}l@{}}{\textbf{Prot2Text-V2}} \\
        SFT & 28.05 & 48.18 & 6.61 \\
        \midrule
        \rowcolor{gray!15}
        \multicolumn{4}{@{}l@{}}{\textbf{Qwen3-8B}} \\
        LLM-only & 0.62 & 21.66 & 4.26 \\
        RAPM & 22.37 & 28.46 & 5.63 \\
        ProtQueSt & 26.69 & 21.00 & 6.05 \\
        QProtFuse & \textbf{32.29} & \textbf{31.53} & \textbf{6.22} \\
        \midrule
        \rowcolor{gray!15}
        \multicolumn{4}{@{}l@{}}{\textbf{DeepSeek-V4-Pro}} \\
        LLM-only & 0.79 & 26.00 & 5.05 \\
        RAPM & 20.66 & 30.93 & 6.26 \\
        ProtQueSt & \textbf{35.43} & 21.20 & 6.51 \\
        QProtFuse & 28.73 & \textbf{30.95} & \textbf{6.68} \\
        \bottomrule
    \end{tabular}
\end{table}

\Cref{tab:generation_results} shows that QProtFuse's retrieval improvements benefit downstream function description. 
On Qwen3-8B~\citep{yang2025qwen3}, QProtFuse outperforms RAPM and ProtQueSt on all three metrics. 
Pairing QProtFuse with DeepSeek-V4-Pro~\citep{xu2026deepseek} yields the highest Judge-F among all compared methods, although ProtQueSt retains the highest Entity-BLEU.
The task-level results and judge scores in Appendix~\ref{sec:appendix_generation_results} further show that QProtFuse improves judge precision over ProtQueSt across all four tasks with both LLMs, suggesting that its retrieved evidence supports more focused, biologically justified descriptions. These results highlight the value of query-guided retrieval for generation quality beyond lexical entity overlap.

\subsection{Ablation and Analysis}
\label{sec:ablation_analysis}

We conduct controlled ablations to isolate the contributions of query-guided fusion and joint training. We further analyze the learned group weights to examine how QProtFuse combines features from different PLM depths.

\begin{figure}[t]
    \centering
    \includegraphics[width=\columnwidth]{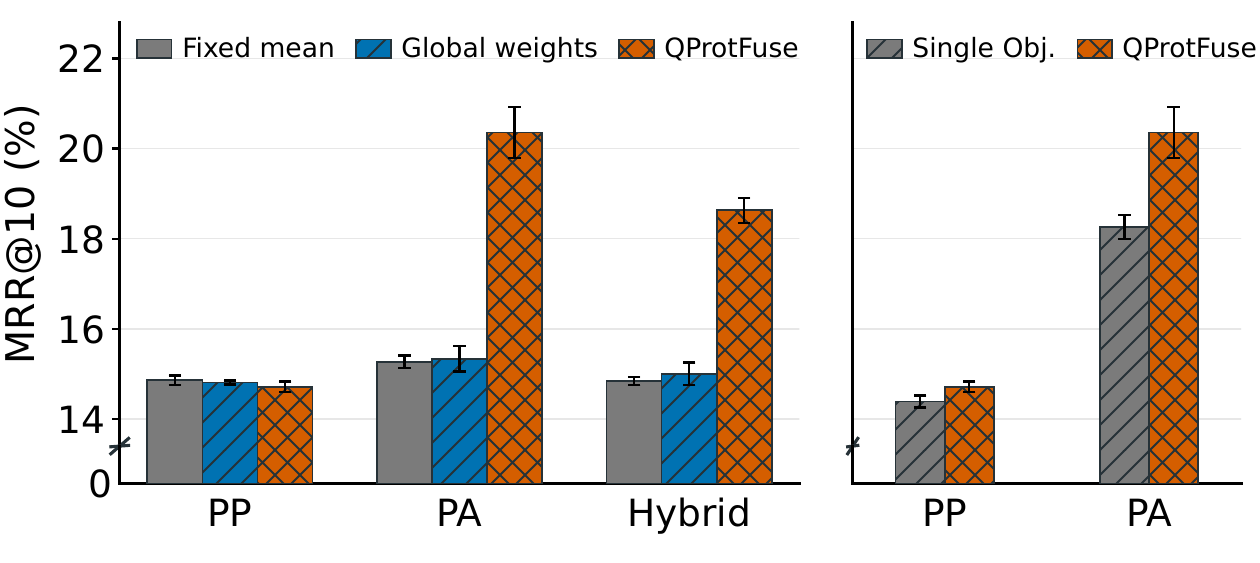}
    \caption{Ablation results. \textbf{Left:} Fusion strategies under joint training. \textit{Fixed mean} and \textit{Global weights} replace the router with uniform all-layer averaging and learned query-independent global group weights, respectively. \textbf{Right:} Joint training versus single PP-only or PA-only training. We report the results over three seeds.}
    \label{fig:ablation}
\end{figure}

\paragraph{Query-guided fusion improves retrieval effectiveness.}
We compare QProtFuse with \textit{fixed mean} and \textit{global weights} variants in \Cref{fig:ablation} (left) to isolate the value of query guidance. 
QProtFuse substantially outperforms both in PA and Hybrid retrieval, indicating better cross-modal alignment between protein and text. 
While PP retrieval performance remains comparable across the three strategies, we speculate that the pooled PLM features already retain sufficient information for protein-to-protein matching, which the final MLP projection can extract under the joint objectives.
Furthermore, learning global weights offers little improvement over fixed mean, highlighting the importance of query-aware layer fusion for effective retrieval.

\begin{figure*}[t]
    \centering
    \includegraphics[width=\textwidth]{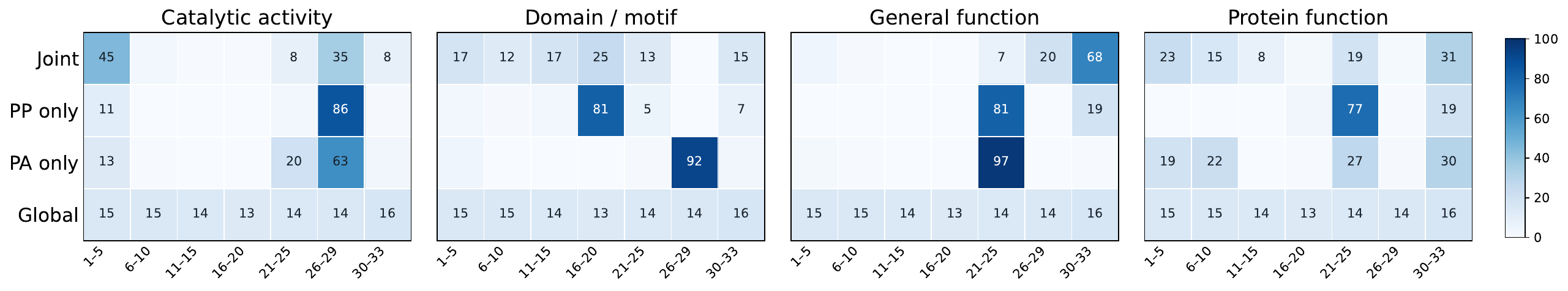}
    \caption{Learned group weights (\%) across four tasks under joint training, PP-only training, PA-only training, and Global weights (\textit{Global}). Columns indicate the ESM layer ranges in each group. Weights are averaged over test queries within each task and across three seeds.}
    \label{fig:group_weights}
\end{figure*}

\begin{figure*}[t]
    \centering
    \includegraphics[width=.92\textwidth]{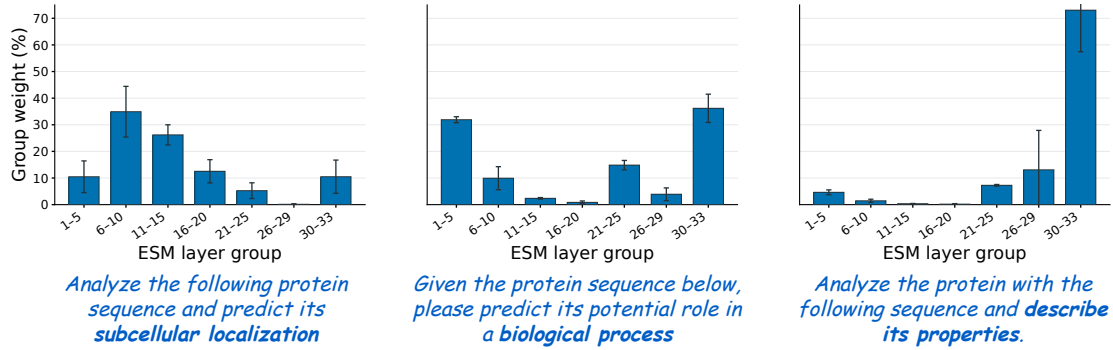}
    \caption{Three examples of query-specific group weights, each with its corresponding query shown below. Bars indicate the mean and standard deviation across three random seeds.}
    \label{fig:weight_cases}
\end{figure*}

\paragraph{The PP and PA objectives reinforce each other.}
As shown in \Cref{fig:ablation} (right), joint training improves both retrieval modes over their respective single-objective models, with a larger gain in PA. This suggests that intra-modal contrastive learning improves protein feature quality and facilitates cross-modal alignment, while cross-modal supervision enriches the functional information captured by these features, benefiting protein retrieval. These findings support the complementary roles of intra-modal and cross-modal contrastive learning~\citep{yang2022vision}.

\paragraph{QProtFuse learns adaptive fusion weights.}
\Cref{fig:group_weights} shows that joint training produces distinct mixtures of early, middle, and late layer features across tasks. 
Catalytic queries combine early (1-5) and late (26-29) layers, general-function queries favor the final group, and domain/motif queries draw more broadly on different depths. 
Single-objective training often exhibits weight collapse toward a single group. We speculate that this limits the expressiveness of the fused features and contributes to weaker retrieval performance. Global weights, in contrast, remain nearly uniform.

\Cref{fig:weight_cases} further reveals distinct fusion patterns for three queries. Subcellular localization emphasizes early-to-middle layers, biological-process prediction combines early and late features, and general property description relies predominantly on the final group.
The contrast between localization and broader property description is qualitatively consistent with prior findings that lower PLM layers capture local structural information, whereas deeper layers encode more complex structural and functional properties~\citep{vig2021bertology,rao2021structure}. 
More examples of query-specific group weights are provided in Appendix Figure~\ref{fig:app_weight_case}. Together, these weights may offer insight into how features from different PLM depths are used to address distinct biological queries.

\section{Conclusion}
\label{sec:conclusion}

In this work, we investigated multi-layer PLM features for protein retrieval and showed that the best-performing layers vary across biological tasks, making conventional final-layer selection and naive all-layer averaging suboptimal.
Building on these findings, we proposed QProtFuse, which adaptively fuses layer features according to the query and jointly optimizes protein and annotation retrieval through intra-modal and cross-modal contrastive learning. 
On Prot-Inst-OOD, QProtFuse achieves the strongest protein and annotation retrieval performance and improves downstream function description. 
Ablations confirm the value of query guidance and the complementary training objectives, while weight analysis reveals distinct fusion patterns for different biological queries. These findings highlight adaptive use of a single PLM's internal knowledge as an effective approach to retrieval-augmented protein function prediction.

\section*{Limitations}

QProtFuse is currently limited by its evaluation scale and its decoupled retrieval-generation design. 
Our current evaluation focuses on Prot-Inst-OOD, which provides a controlled setting for studying query-guided protein retrieval and its effect on function description. Scaling to larger protein resources, such as UniProt~\citep{uniprot2018uniprot}, could provide more diverse biological knowledge and support further performance gains.
In addition, our method focuses on improving retrieval while keeping the LLM frozen. Generated responses are not subsequently refined or verified against the retrieved evidence. This design isolates the contribution of retrieval and leaves room for further improvements through evidence-aware generation, response verification, and joint optimization of retrieval and generation.

\section*{Ethical Considerations}

\paragraph{Potential Misuse.}
Query-guided feature fusion can support protein research, but it also carries potential risks of misuse. Malicious queries could direct the system to identify proteins relevant to harmful biological applications. Responsible deployment should therefore incorporate query screening, appropriate access controls, and human review of potentially sensitive requests.

\paragraph{AI Assistance.}
We used OpenAI Codex to assist with code development and manuscript drafting. For each manuscript section, the authors provided an outline specifying the logical structure, from which the AI produced an initial draft. The authors then revised and polished these drafts. The authors take responsibility for the final manuscript, implementation, and reported results.

\bibliography{custom}

\appendix
\clearpage
\section{Single-Layer Features versus Uniform Averaging}
\label{sec:appendix_layer_comparison}

\Cref{tab:single_layer_comparison} compares uniform averaging with the best single layer selected post hoc by MRR@10 for each task. The selected layers vary across tasks and generally outperform uniform averaging, motivating adaptive fusion over a fixed average.

\begin{table}[htbp]
    \centering
    \fontsize{10pt}{11pt}\selectfont
    \setlength{\tabcolsep}{3pt}
    \caption{ESM2-650M single-layer and all-layer mean retrieval results on Prot-Inst-OOD (\%). Single layers are selected by task-specific MRR@10. Both use test-to-train retrieval with instruction cosine similarity $\geq 0.65$}
    \label{tab:single_layer_comparison}
    \begin{tabular}{@{}lcccc@{}}
        \toprule
        Task & Layer & MRR@10 & R@1 & R@10 \\
        \midrule
        \rowcolor{gray!15}
        \multicolumn{5}{@{}l@{}}{\bf Best single layer} \\
        Catalytic activity & 17 & 17.46 & 13.39 & 25.97 \\
        Domain/motif & 15 & 21.40 & 13.87 & 38.51 \\
        General function & 29 & 5.75 & 3.26 & 11.61 \\
        Protein function & 30 & 12.66 & 8.00 & 23.62 \\
        \midrule
        \rowcolor{gray!15}
        \multicolumn{5}{@{}l@{}}{\bf All-layer mean} \\
        Catalytic activity & 1--33 & 16.89 & 12.78 & 25.87 \\
        Domain/motif & 1--33 & 20.45 & 12.88 & 37.85 \\
        General function & 1--33 & 5.65 & 3.16 & 11.43 \\
        Protein function & 1--33 & 12.82 & 8.26 & 23.36 \\
        \bottomrule
    \end{tabular}
\end{table}

\section{Prot-Inst-OOD Dataset}
\label{sec:appendix_dataset}

Prot-Inst-OOD is constructed by repartitioning the protein understanding tasks in Mol-Instructions~\citep{fang2024mol}. 
Its four tasks cover protein functions, including molecular roles, biological processes, and localization; catalytic activities and reactions; domains and motifs; and general functional descriptions. 
Protein sequences are clustered with MMseqs2, and clusters are randomly assigned to training and test sets at an initial ratio of 80:20. 
To reduce annotation leakage, each test protein is matched to its nearest training protein with MMseqs2. Test samples sharing the same label as their retrieved counterpart are moved to the training set, and this filtering step is repeated twice. 
\Cref{tab:dataset_statistics} summarizes the sample counts and the number of distinct instructions for each task.

\begin{table}[h]
    \centering
    \fontsize{10pt}{11pt}\selectfont
    \setlength{\tabcolsep}{4pt}
    \caption{Prot-Inst-OOD statistics. The final column reports the number of distinct instruction types.}
    \label{tab:dataset_statistics}
    \begin{tabular}{@{}lrrr@{}}
        \toprule
        Task & Train & Test & Instructions \\
        \midrule
        Protein function & 108,696 & 5,487 & 38 \\
        Catalytic activity & 51,187 & 1,987 & 5 \\
        Domain/motif & 42,368 & 2,732 & 10 \\
        General function & 82,275 & 4,297 & 7 \\
        \midrule
        Total & 284,526 & 14,503 & 60 \\
        \bottomrule
    \end{tabular}
\end{table}

\section{Evaluation Metrics}
\label{sec:appendix_generation_metrics}

\paragraph{Retrieval metrics.}
A retrieved record counts as a hit only when its bio-entity annotation exactly matches the query protein's ground-truth annotation. We report Recall@1 and Recall@10, defined as the fractions of queries with at least one hit among the top 1 and top 10 results, respectively. MRR@10 additionally measures how highly the first hit is ranked:
\begin{equation}
    \operatorname{MRR}@10 = \frac{1}{N}\sum_{i=1}^{N}
    \begin{cases}
        1/r_i, & r_i\leq 10, \\
        0, & \text{otherwise},
    \end{cases}
\end{equation}
where $N$ is the number of evaluated queries and $r_i$ is the rank of the first hit for query $i$. All three metrics are computed per task and macro-averaged across the four tasks.

\paragraph{ROUGE-L.}
ROUGE-L~\citep{lin2004rouge} measures overlap through the longest common subsequence (LCS) between a prediction $\hat{d}$ and reference $d$. We report its F1 score:
\begin{equation}
    \operatorname{ROUGE\text{-}L}_{\mathrm{F1}}
    = \frac{2\operatorname{LCS}(\hat{d},d)}{|\hat{d}|+|d|},
\end{equation}
where lengths and LCS are computed over tokens. We use the Python package \texttt{rouge-score} version 0.1.2 with \texttt{RougeScorer(["rougeL"],\allowbreak{} use\_stemmer=False)}.

\paragraph{Entity-BLEU.}
Entity-BLEU~\citep{wu-etal-2025-rethinking-text} measures biological content overlap rather than overlap in generic response wording. Biological entities are first extracted from predictions and references using RAPM's vocabulary derived from InterPro, EC-GO, and Mol-Instructions metadata. Standard BLEU~\citep{papineni2002bleu} is then computed on the extracted sequences:
\begin{equation}
    \operatorname{E\text{-}BLEU}_{k}
    = \operatorname{BP}\cdot \exp\!\left(\frac{1}{k}\sum_{n=1}^{k}\log p_n\right),
    \label{eq:entity_bleu}
\end{equation}
where $p_n$ is the corpus-level clipped $n$-gram precision of the extracted sequences, and $\operatorname{BP}=\min\{1,\exp(1-L_{\mathrm{ref}}/L_{\mathrm{pred}})\}$ penalizes predictions shorter than their references. Here, $L_{\mathrm{pred}}$ and $L_{\mathrm{ref}}$ are the total extracted sequence lengths. We use $k=2$, combining unigram and bigram precision, and report scores as percentages.

\begin{figure}[t]
    \centering
    \includegraphics[width=\linewidth]{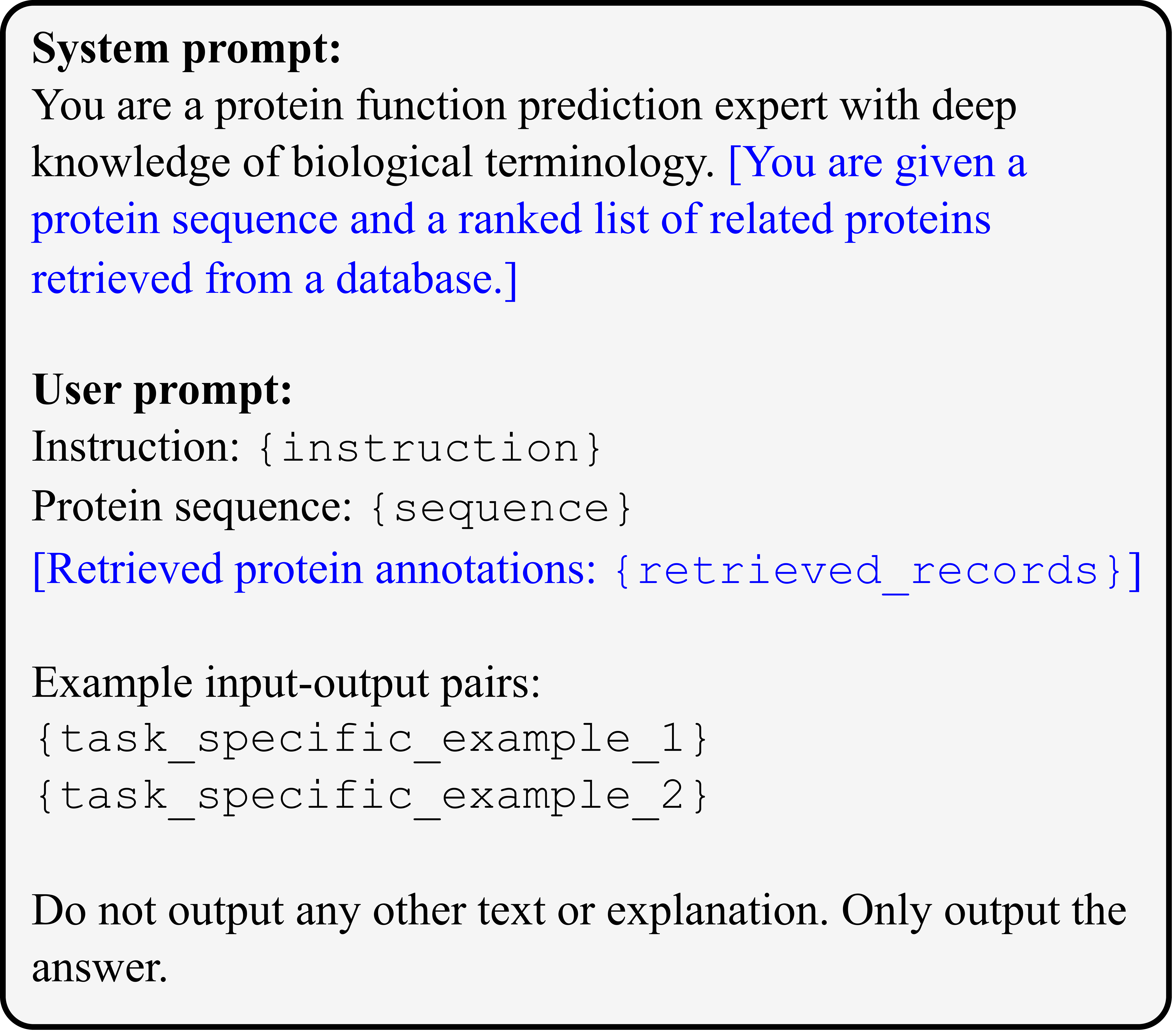}
    \caption{Generation prompt. Blue text is included only for RAG.}
    \label{fig:generation_prompt}
\end{figure}

\paragraph{LLM-as-a-judge.}
Given the instruction, reference description, reference bio-entities, and prediction, the judge assesses four axes: \textit{recall} of reference entities, \textit{precision} of supported or reasonably inferred claims, \textit{specificity} of biological content, and \textit{plausibility} without contradictions or fabrications. Each active axis receives a score in $\{0,3,5,7,10\}$. The final per-sample score and the macro-average are
\begin{align}
    &J_i = \operatorname{round}\!\left(\frac{1}{|\mathcal{A}_i|}
    \sum_{a\in\mathcal{A}_i}g_{i,a}\right), \\
    &J_{\mathrm{macro}} = \frac{1}{4}\sum_{t=1}^{4}
    \frac{1}{N_t}\sum_{i\in\mathcal{I}_t}J_i,
\end{align}
where $g_{i,a}$ is the score for axis $a$, $\mathcal{A}_i$ excludes axes marked inactive ($-1$), and $\mathcal{I}_t$ contains the $N_t$ evaluated samples for task $t$. The final score ranges from 0 to 10. We refer readers to ProtQueSt for the complete prompt and rubric.

\begin{table*}[htbp]
    \centering
    \fontsize{9pt}{10pt}\selectfont
    \setlength{\tabcolsep}{3pt}
    \caption{Per-task generation results on Prot-Inst-OOD. E-BLEU denotes Entity-BLEU-2 and RG-L denotes ROUGE-L (\%); Judge-F is on a 0--10 scale. Bold denotes column-wise maxima within each general-purpose LLM block.}
    \label{tab:generation_per_task}
    \begin{tabular}{@{}l*{12}{c}@{}}
        \toprule
        & \multicolumn{3}{c}{Protein function} & \multicolumn{3}{c}{General function} & \multicolumn{3}{c}{Domain/motif} & \multicolumn{3}{c}{Catalytic activity} \\
        \cmidrule(lr){2-4}\cmidrule(lr){5-7}\cmidrule(lr){8-10}\cmidrule(lr){11-13}
        Method & E-BLEU & RG-L & Judge-F & E-BLEU & RG-L & Judge-F & E-BLEU & RG-L & Judge-F & E-BLEU & RG-L & Judge-F \\
        \midrule
        \rowcolor{gray!15}
        \multicolumn{13}{@{}l@{}}{\textbf{Prot2Text-V2}} \\
        SFT & 42.39 & 50.31 & 7.22 & 16.82 & 42.52 & 5.75 & 26.27 & 45.98 & 7.23 & 26.74 & 53.92 & 6.25 \\
        \midrule
        \rowcolor{gray!15}
        \multicolumn{13}{@{}l@{}}{\textbf{Qwen3-8B}} \\
        LLM-only & 0.04 & 18.29 & 4.00 & 0.02 & 18.30 & 3.57 & 2.40 & 16.36 & 4.48 & 0.00 & 33.68 & 4.98 \\
        RAPM & 29.10 & 25.32 & 6.44 & 7.39 & 25.69 & 5.03 & 38.24 & 23.79 & 5.42 & 14.73 & 39.05 & 5.62 \\
        ProtQueSt & \textbf{36.25} & 22.29 & 6.63 & \textbf{18.79} & 24.01 & \textbf{5.43} & 19.34 & 8.66 & 5.73 & 32.37 & 29.05 & \textbf{6.38} \\
        QProtFuse & 34.99 & \textbf{27.18} & \textbf{6.87} & 10.28 & \textbf{29.33} & 5.30 & \textbf{48.61} & \textbf{29.25} & \textbf{6.62} & \textbf{35.27} & \textbf{40.36} & 6.09 \\
        \midrule
        \rowcolor{gray!15}
        \multicolumn{13}{@{}l@{}}{\textbf{DeepSeek-V4-Pro}} \\
        LLM-only & 1.86 & 25.47 & 5.02 & 0.12 & 26.62 & 4.23 & 0.01 & 13.30 & 5.76 & 1.17 & \textbf{38.63} & 5.20 \\
        RAPM & 33.92 & 30.40 & 6.93 & 5.29 & 32.56 & 5.58 & 27.23 & \textbf{23.10} & 6.75 & 16.21 & 37.65 & 5.77 \\
        ProtQueSt & \textbf{48.25} & 22.11 & 7.00 & \textbf{18.66} & 24.58 & \textbf{5.72} & 37.56 & 11.53 & 6.65 & \textbf{37.26} & 26.60 & \textbf{6.68} \\
        QProtFuse & 36.34 & \textbf{30.77} & \textbf{7.17} & 6.05 & \textbf{33.45} & 5.60 & \textbf{40.57} & 21.74 & \textbf{7.63} & 31.97 & 37.83 & 6.31 \\
        \bottomrule
    \end{tabular}
\end{table*}

\begin{table*}[h]
    \centering
    \fontsize{9pt}{10pt}\selectfont
    \setlength{\tabcolsep}{3pt}
    \caption{Per-task LLM-as-a-judge scores (0--10). Rec., Prec., Spec., and Plaus. denote recall, precision, specificity, and plausibility, respectively. Bold denotes column-wise maxima within each general-purpose LLM block.}
    \label{tab:generation_judge_axes}
    \begin{tabular}{@{}l*{16}{c}@{}}
        \toprule
        & \multicolumn{4}{c}{Protein function} & \multicolumn{4}{c}{General function} & \multicolumn{4}{c}{Domain/motif} & \multicolumn{4}{c}{Catalytic activity} \\
        \cmidrule(lr){2-5}\cmidrule(lr){6-9}\cmidrule(lr){10-13}\cmidrule(lr){14-17}
        Method & Rec. & Prec. & Spec. & Plaus. & Rec. & Prec. & Spec. & Plaus. & Rec. & Prec. & Spec. & Plaus. & Rec. & Prec. & Spec. & Plaus. \\
        \midrule
        \rowcolor{gray!15}
        \multicolumn{17}{@{}l@{}}{\textbf{Prot2Text-V2}} \\
        SFT & 5.26 & 7.37 & 8.63 & 7.36 & 3.04 & 6.70 & 8.17 & 4.93 & 4.04 & 7.90 & 8.80 & 7.87 & 2.30 & 7.57 & 9.69 & 4.82 \\
        \midrule
        \rowcolor{gray!15}
        \multicolumn{17}{@{}l@{}}{\textbf{Qwen3-8B}} \\
        LLM-only & 0.51 & 4.69 & 7.45 & 2.91 & 0.04 & 4.95 & 4.31 & 4.60 & 0.56 & 2.70 & 7.43 & 6.71 & 0.00 & 6.99 & \textbf{9.81} & 2.60 \\
        RAPM & 5.32 & 5.31 & 8.74 & 6.07 & 3.49 & 4.54 & 6.97 & 4.82 & 4.08 & 2.22 & 8.41 & 6.64 & 1.28 & 7.14 & 9.69 & 3.83 \\
        ProtQueSt & 5.63 & 5.13 & \textbf{9.17} & 6.22 & \textbf{4.13} & 4.29 & \textbf{8.76} & 4.42 & 3.76 & 3.46 & \textbf{8.86} & 6.43 & \textbf{3.92} & 6.67 & 9.44 & \textbf{4.97} \\
        QProtFuse & \textbf{5.92} & \textbf{5.74} & 8.75 & \textbf{6.66} & 3.61 & \textbf{5.05} & 7.27 & \textbf{4.97} & \textbf{5.88} & \textbf{4.52} & 8.45 & \textbf{7.07} & 2.44 & \textbf{7.27} & 9.62 & 4.46 \\
        \midrule
        \rowcolor{gray!15}
        \multicolumn{17}{@{}l@{}}{\textbf{DeepSeek-V4-Pro}} \\
        LLM-only & 2.00 & 5.71 & 6.79 & 5.30 & 0.70 & 5.41 & 5.76 & 4.81 & 0.96 & 6.84 & 7.90 & 6.90 & 0.58 & 7.18 & 9.24 & 3.37 \\
        RAPM & 5.67 & 6.23 & 8.91 & 6.59 & 3.72 & 5.69 & 7.24 & \textbf{5.37} & 3.96 & 6.59 & 8.65 & 7.28 & 1.80 & 7.37 & \textbf{9.56} & 4.02 \\
        ProtQueSt & \textbf{6.65} & 5.17 & \textbf{9.34} & 6.44 & \textbf{4.32} & 4.63 & \textbf{8.58} & 5.00 & 4.98 & 5.20 & \textbf{9.01} & 6.80 & \textbf{4.76} & 6.68 & 9.27 & \textbf{5.57} \\
        QProtFuse & 6.04 & \textbf{6.53} & 8.89 & \textbf{6.83} & 3.49 & \textbf{5.96} & 7.37 & 5.26 & \textbf{5.58} & \textbf{7.59} & 8.96 & \textbf{7.86} & 2.70 & \textbf{7.67} & 9.46 & 4.97 \\
        \bottomrule
    \end{tabular}
\end{table*}

We adopt ProtQueSt's prompt and scoring rubric~\citep{ma2026protquest}, using Qwen3.8-27B~\cite{qwen38} in place of Claude as the judge. We set the temperature to $0.0$ and the total generation budget to 5,024 tokens, including a 4,000-token thinking budget. Following ProtQueSt, we randomly sample 500 predictions per task, yielding 2,000 evaluated samples per method.

\section{Generation Details}
\label{sec:appendix_generation_prompts}

\paragraph{Prompt.} \Cref{fig:generation_prompt} shows our generation template. All settings use the same two randomly selected examples per task, each formatted as ``\{\texttt{Instruction: ...}, \texttt{Protein Sequence: ...}, \texttt{Description: ...}\}''. Example sequences longer than 80 amino acids are truncated to their first 80 residues followed by `\texttt{...}'.
The placeholder \texttt{\{retrieved\_records\}} is filled with a list of ten retrieved samples organized as ``\texttt{[\{'db\_label': ..., 'protein\_similarity': ..., 'protein\_text\_alignment': ...\}, ...]}''.
For RAPM and ProtQueSt, we use their respective published prompts.

\paragraph{Inference and sampling.}
We run Qwen3-8B locally with vLLM~\citep{kwon2023efficient} and access DeepSeek-V4-Pro through its online API, using the \texttt{deepseek-v4-pro-0813} version. Both models use a temperature of $1.0$, a maximum output length of 512 tokens, and disabled thinking.

\begin{figure*}[t]
    \centering
    \includegraphics[width=\linewidth]{figs/appdx_weight_cases.pdf}
    \caption{More examples of query-specific group weights, each with its corresponding query shown below. }
    \label{fig:app_weight_case}
\end{figure*}

\section{Training Configuration}
\label{sec:appendix_training}

\paragraph{QProtFuse.} Our model has 5.13M trainable parameters. We train it for 30 epochs with three random seeds (20260929, 20260930, and 20261001) using mixed bfloat16 precision on four NVIDIA H200 GPUs. The batch size is 256 per GPU, yielding a global batch size of 1,024. Each training run takes approximately 0.36 GPU hours, excluding the precomputation of ESM features, which are cached locally. We use AdamW with learning rates of $10^{-4}$ for the router and $10^{-3}$ for the remaining modules, and a weight decay of $10^{-4}$. All reported results use the last checkpoint.

\paragraph{Baselines.}
For Prot2Text-V2, only the projection MLP and the LLM LoRA adapters are trained, with LoRA rank 32, scaling factor 64, and dropout $0.1$. Training runs for up to 24 epochs on 16 NVIDIA H200 GPUs using mixed bfloat16 precision. A per-GPU batch size of 4 and four gradient-accumulation steps yield an effective batch size of 256. We use AdamW with a learning rate of $2\times10^{-4}$, weight decay of $0.01$, cosine learning-rate decay, and a random seed of 42.
For ProtQueSt, whose code was unavailable for our experiments, we independently implement and train the model following the configuration disclosed in its paper.

\section{Detailed Generation Results}
\label{sec:appendix_generation_results}

\Cref{tab:generation_per_task} reports generation metrics for each task, and \Cref{tab:generation_judge_axes} breaks down the four judge dimensions. QProtFuse achieves higher judge precision than ProtQueSt in every task with both generators.

\end{document}